\documentclass[sigconf, anonymous=false]{acmart}

\usepackage{booktabs}
\usepackage{graphicx}
\usepackage{amssymb}

\usepackage{bm}
\renewcommand*{\vec}[1]{\bm{#1}}
\usepackage{mathtools}
\usepackage{bbm}
\usepackage{multirow}
\usepackage{graphicx}
\usepackage[algoruled,vlined]{algorithm2e}
\usepackage{cleveref}

\usepackage{subcaption}

\usepackage{url}
\def\BibTeX{{\rm B\kern-.05em{\sc i\kern-.025em b}\kern-.08emT\kern-.1667em\lower.7ex\hbox{E}\kern-.125emX}}
    
\copyrightyear{2020}
\acmYear{2020} 
\setcopyright{iw3c2w3}
\acmConference[WWW '20]{Proceedings of The Web Conference 2020}{April 20--24, 2020}{Taipei, Taiwan} 
\acmBooktitle{Proceedings of The Web Conference 2020 (WWW '20), April 20--24, 2020, Taipei, Taiwan}
\acmPrice{}
\acmDOI{10.1145/3366423.3380274}
\acmISBN{978-1-4503-7023-3/20/04}

\begin{document}

\title{Discovering Strategic Behaviors for Collaborative Content-Production in Social Networks}

\author{Yuxin Xiao, Adit Krishnan, Hari Sundaram}
\affiliation{%
    \institution{University of Illinois at Urbana-Champaign}
    \institution{\{yuxinx2, aditk2, hs1\}@illinois.edu}
}

\renewcommand{\shortauthors}{Y. Xiao et al.}

\begin{abstract}
Some social networks provide explicit mechanisms to allocate social rewards such as reputation based on users' actions, while the mechanism is more opaque in other networks. Nonetheless, there are always individuals who obtain greater rewards and reputation than their peers. An intuitive yet important question to ask is whether these successful users employ strategic behaviors to become influential. It might appear that the influencers "have gamed the system." However, it remains difficult to conclude the rationality of their actions due to factors like the combinatorial strategy space, inability to determine payoffs, and resource limitations faced by individuals. The challenging nature of this question has drawn attention from both the theory and data mining communities.
Therefore, in this paper, we are motivated to investigate if resource-limited individuals discover strategic behaviors associated with high payoffs when producing collaborative/interactive content in social networks. We propose a novel framework of Dynamic Dual Attention Networks (DDAN) which models individuals' content production strategies through a generative process, under the influence of social interactions involved in the process. Extensive experimental results illustrate the model's effectiveness in user behavior modeling. We make three strong empirical findings: (1) Different strategies give rise to different social payoffs; (2) The best performing individuals exhibit stability in their preference over the discovered strategies, which indicates the emergence of strategic behavior; and (3) The stability of a user's preference is correlated with high payoffs. 
\end{abstract}


\maketitle

\section{Introduction} \label{sec:1}
This paper examines if individuals can successfully discover strategies with high payoffs in social networks. In seminal work,~\citet{simon1972theories} introduced the idea of bounded rationality---that human beings use limited resources to make decisions. In more recent work, Gigerenzer et al.~\cite{gigerenzer1996reasoning, gigerenzer2011heuristic} argued that human beings used heuristics to make decisions whose quality matched that of rational agents. 
Online social networks typically have an explicit mechanism that allocates rewards (usually points) that vary with users' behaviors; for example, the right answer on a community question-answer website like StackOverflow (\url{https://www.stackoverflow.com}) may earn the individual who posted the answer reputation points.

In some other networks, the mechanism is more opaque. For example, on Twitter, the inclusion of a celebrity's twitter handle on your tweet causes your tweet to appear on their timeline, increasing your visibility. Perhaps over time, this improved visibility results in one having more followers, and that increased visibility may cause one to become an influencer---advertisers may reach out to market their products. In a different example, an assistant professor needs to decide where she should publish her current work. Should she submit the paper to a high prestige conference with a lower probability of acceptance or a lower-tier conference with a higher probability? The former strategy yields greater visibility, but with a lower acceptance rate. If the conference rejects her paper, she may need to wait out a year. Some individuals on websites with explicit mechanisms have many reputation points; others in networks such as Twitter, are influencers---did these individuals employ strategic behaviors to gain points or to become influential?

At first glance, it might appear that individuals who do well, ``have gamed the system'' as it were, and the rest have not figured out the mechanism. However, for games with opaque mechanisms, the strategy space is unclear; and for games with explicit mechanisms (e.g., StackOverflow), the payoffs for a particular action are still unknown. It is not straightforward to conclude that the winners of these social networks (e.g., influencers on Twitter) are rational in the classic sense (i.e., maximize expected utility) due to several reasons. The combinatorial strategy space (e.g., on StackOverflow, which question to answer, when, answer length, readability, etc.); the inability to determine the payoffs (e.g., reputation points) for any given strategy; and the fact that individuals do not have unlimited resources to determine their best response. Instead, the best we can conclude is that the best players are \textit{differentially rational}---that is, when compared to their peers, they have a better understanding of the correlation between a strategy and its payoff.

The theoretical Computer Science community has paid attention to games of incomplete information~\cite{hartline2015no,lykouris2016learning,feng2018learning} with dynamic populations (as in behavior in online auctions). For example, a key result from~\citet{lykouris2016learning} is that when agents play repeated games with strategies that guarantee low-adaptive regret, high social welfare is ensured. One of the challenges with theoretical work is that it is unclear if, in practice, individuals can find successful strategies. Most existing social modeling tasks target on discovering people's interests from textual contents on social media \cite{zhu2017next, bhattacharya2014inferring, li2008tag, qiu2013not} or tracing the propagation of social influence along social networks \cite{papagelis2011individual, barbieri2013topic, tang2009social}. A few do consider the latent strategies adopted by people in social networks; however, they either ignore contextual information \cite{dong2014inferring}, or do not further examine the impact of those strategies \cite{xu2012modeling}. 
Thus we are motivated to ask a simple question: 
\begin{quote}
    Can individuals with limited resources discover content production strategies with high payoffs in social networks?
\end{quote}
To operationalize our question, we analyze the preference order over strategies. In particular, we ask two questions: first, does the preference order among strategies for authors stabilize over time, indicating the emergence of strategic behavior? Second, if the preference order is stable, does the preference order maximize utility? Notice that preference order stability does not imply high payoffs; the stability may arise due to other factors such as social norms.

We wish to answer this question through an analysis of empirical data from a social network. An empirical analysis is non-trivial: while we may observe a particular outcome (e.g., which paper to cite; which celebrity's handle to mention; the topic of the message that we post on social media), as well as be able to compute the reward, \textit{we do not observe the strategic considerations} underlying the action.

\noindent\textbf{Our technical insight:} to model the observed behavior as a generative process. That is, a strategic decision changes the posterior distribution over the action space. We assume that while the set of strategies is common to all, each individual adopts a mixed-strategy over the set of different strategies. In other words, the distribution over the set of strategies is private to each individual. 
To model individual behavior, we propose conceptualizing content production as a bipartite graph where the nodes include individuals and contents, and where content may have multiple authors. Thus, the strategy to produce a piece of content (e.g., author a paper in an academic social network, post blogs online) depends on the strategy distributions of its authors; and the co-authors may influence the strategy distribution of an author. We identify an elegant dual attention neural architecture motivated by~\citet{velivckovic2017graph} to model individual behavior. Then, we compare our results with a counterfactual condition: the inferred strategic behavior of an idealized expected-utility maximizer.
We summarize our contributions as follows:

\begin{description}
    \item[Coupling authors and content:] We propose a novel Dynamic Dual Attention Network (DDAN) to jointly model the role of the authors in the determination of content production strategy, and how co-authoring content influences authors' content production strategy. The DDAN helps discover the author's strategy. In contrast, past works either focus on theoretical concepts (e.g., ~\cite{hartline2015no,lykouris2016learning}) or do not attempt to identify strategic behaviors from data. In the dynamic dual attention mechanism, the content strategy depends on the strategies of all of its authors. Conversely, the strategy of an author depends on her prior production strategy as well as the strategies of all the content that she played a role in producing at the current moment. Extensive experiments show that our framework models user behaviors well.
    \item[Strong experimental findings:] We have strong qualitative findings. First, we show that different strategies result in different payoffs. Second, we show through rank correlation, that the authors with the top $10\%$ normalized utility exhibit stability in their preference orders. Furthermore, a majority of authors do not discover the correlation between strategies and payoffs. Third, we show that the stability of preference is correlated with high payoffs. 
\end{description}

\noindent\textbf{The significance of this work:} to the best of our knowledge, this is the first attempt to identify strategic behaviors from empirical data formally.
We organize the rest of the paper as follows. In the next section, we introduce the problem. In~\Cref{sec:3}, we show how to model strategic behavior, including identifying our assumptions, and introducing our Dynamic Dual Attention Network formalism. In~\Cref{sec:4}, we propose a model for rational behavior, to serve as an idealized baseline. Then, in~\Cref{sec:5}, we present experiments on an academic dataset, including specification of the strategic space.~\Cref{sec:6} highlights the qualitative findings. Then, we discuss challenges and limitations in~\Cref{sec:discussion} followed by a discussion of related work in~\Cref{sec:7}. We conclude in~\Cref{sec:8}.

\begin{figure}[t]
\includegraphics[width=\columnwidth]{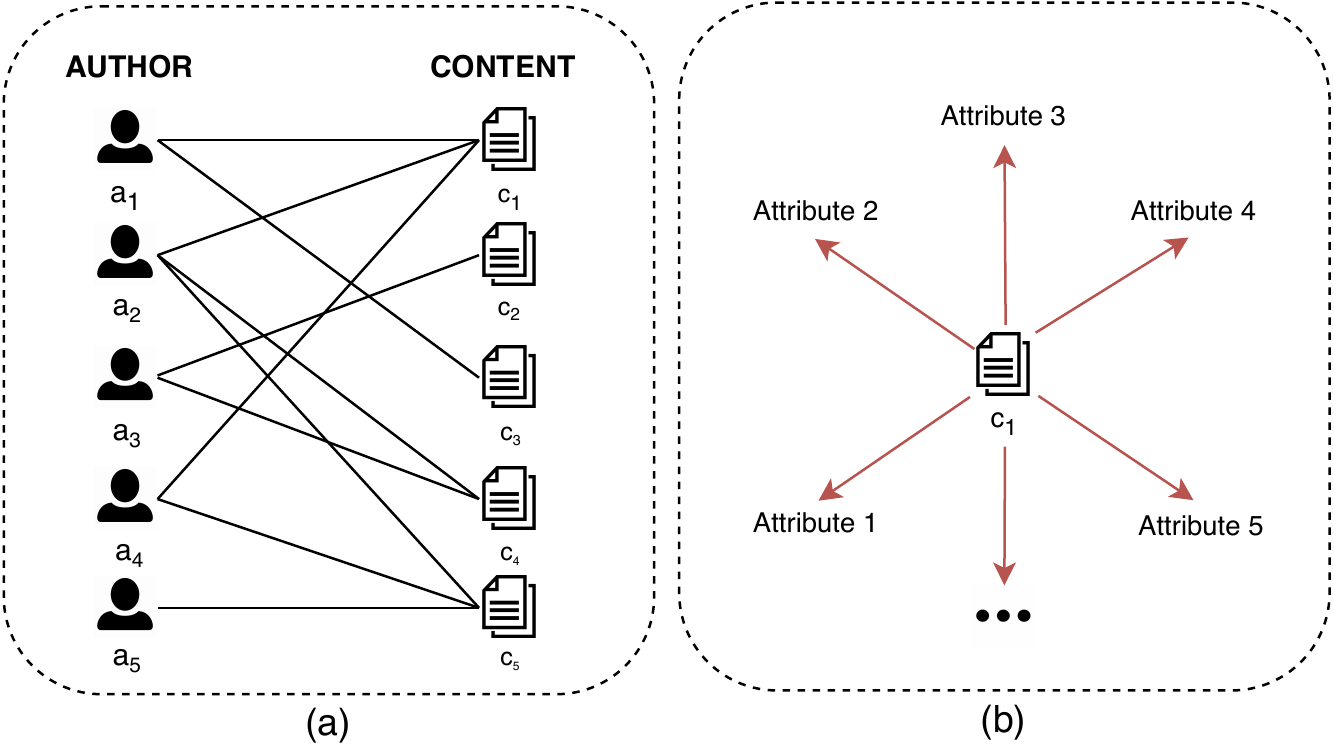}
  \caption{(a) Authors and contents form a bipartite graph where the edges between them indicating that a group of authors work together to produce contents. (b) Each content has multiple attributes, including but not limited to, time and venue of publication, textual topic, and links to other entities.} \label{fig:section2}
\end{figure}
\section{Problem Formulation} \label{sec:2}
We first present an informal description before presenting the problem in detail.
\subsection{Informal Problem Description} \label{sub:2.1}
Consider a general scenario, where a set of individuals $\mathcal{A}$ work together to author a piece of content $c$. This content could be a blog post, an academic paper, or when a group attempts to answer a question on a community question-answer forum such as StackExchange. 
We can associate attributes with the created content $c$, including venue of publication; time of publication; content topic; links to other entities including contents (e.g., citations to other academic papers, links to other blog posts) and authors (e.g., tweets can include mentions
of other individuals on Twitter).

We can associate a time-varying utility $\mu_c(k)$ to each content $c$ published at time $t$, where we evaluate the utility after $k$ time units (i.e., at time $t+k$): academic papers receive citations; blog posts receive in-links; answers on a community question answer forum receive up or down votes (the forum will transform the votes to reputation points). In general, the relationship between a particular choice of attributes for the content (e.g., topic) and its utility is non-trivial to assess. Not only is the relationship non-deterministic, but also, the payoffs are distant.

Authors have to make decisions about content attributes, including, for example, the topic, the publication venue, links to other entities. We identify four challenges. First, the attribute cardinality is large. For example, for an author to identify a paper to cite, she needs to consider \textit{the entire set of past published papers}. Instead, she may have \textit{private} strategies, including picking papers that are highly cited, to bias her attribute selection. Second, the mapping between attribute choice and utility is non-deterministic, with a distant payoff. Third, while she can observe the choice of attributes (e.g., content topic), made by her peers, and their payoffs (e.g., citations), \textit{she cannot observe the strategic consideration behind that choice}. Finally, authors are resource-limited (e.g., limited time, attention), limiting their ability to determine co-variation between their choices and payoffs. Thus, we can ask:
\begin{quote}
Are resource-limited individuals in social networks able to discover content-production strategies that yield high payoffs?
\end{quote}
\subsection{Data Model} \label{sec:2.2}
Now, we develop the data model for the problem. Let $\mathcal{A}$ denote the set of individuals who produce content on a social network, and let $\mathcal{C}$ denote the set of content. Authors may either collaborate or work alone to author content $c \in \mathcal{C}$. Thus, we can construct an undirected bipartite graph $G_{a,c} = (\mathcal{V}, \mathcal{E})$, where $\mathcal{V} = \mathcal{A} \cup \mathcal{C}$, $\mathcal{E} = \{(a,c) \: | \: a \in \mathcal{A}, \: c \in \mathcal{C}, \: a \text{ is an author of } c \}$, to compactly represent content production~(\Cref{fig:section2}  (a)).~\Cref{fig:section2} (b) illustrates that each content $c \in \mathcal{C}$ may have multiple attributes.

The author picks the attribute values strategically. Let every author use the \textit{same} strategy space $\mathcal{S}$, where $|\mathcal{S}|=m$. However, each author randomizes over the $m$ strategies independently. That is, for each author $a$, we associate a probability distribution $\vec{D}_a$ over the $m$ strategies, \textit{private to each individual}, from which she draws her strategy to determine the attributes for $c$. For example, if an author wishes to determine which papers to cite, her strategies could include picking papers uniformly at random from past papers, as well as picking papers based on citation count. More formally, each choice of strategy $S \in \mathcal{S}$ affects the posterior distribution of attribute values. When a group of authors collaborate, we assume that they negotiate and develop a consensus strategy.  We use $\vec{D}_a(t)$ to represent $a$'s strategy distribution at time $t$. We assume that the author's past strategy $\vec{D}_a(t-1)$ and her co-authors' strategy at time $t$ influence $\vec{D}_a(t)$.

Assume that a content $c$ authored by $a$ at time $t$ receives a utility $\mu_c(k)$ after $k$ time units. 
In general, at time $t$, an author $a$ creates a set of content $\mathcal{C}_a(t)$. Some content $c \in \mathcal{C}_a(t)$ may have co-authors. Thus the total utility for author $a$ for authoring these contents $\mathcal{C}_a(t)$ at time $t$,  $k$ time units after publication,  is:
\begin{equation}
  \mu_a(t, k) = \sum_{c \in \mathcal{C}_a(t)} \mu_c(k) \cdot r( a \mid c), \label{eq:author_contribution}
\end{equation}
where, $r( a \mid c) \in [0,1]$ is the attribution of author $a$'s role in creating content $c$. Notice for each $c \in C_a(t)$, the co-author group may be different.
Each author $a$ has a \textit{private} preference between a pair of strategies $S_i, S_j$. We denote $S_i \succ_a S_j$ (or $S_i$ dominates $S_j$) if the probability of picking $S_i$ is greater than $S_j$ and $S_i \sim_a S_j$ if $a$ is indifferent between the two. We ask two questions:
\begin{quote}
  First, does the preference order among strategies for authors stabilize over time, indicating the emergence of strategic behaviors? Second, if the preference order is stable, does the preference order maximize utility?  
\end{quote}
Notice that the emergence of stable preferences by itself does not imply that the author is maximizing utility, since social norms may cause preference stability. We will compare the strategic behaviors of social network participants against myopic rational agents that maximize expected utility. 
\begin{table}[t]
    \centering
    \begin{tabular}{ll}
        \toprule
            Notation & Description \\
        \midrule
            $G(t)$ & Snapshot of the author-content graph at time $t$  \\
            $\mathcal{C}_a(t)$ & Set of contents created by $a$ at time $t$ \\
            $\widetilde{\mathcal{A}}'(t)$ & Set of authors with over 5 active contents at time $t$ \\    
            $\Vec{h}_a(t)$ & Embedding vector of $a$ at time $t$ \\
            $\Vec{F}_a(t)$ & Field vector of $a$ at time $t$ \\
            $\vec{D}_a(t)$ & Strategy distribution of $a$ at time $t$ \\
            $r( a \mid c)$ & $a$'s contribution to $c$'s strategy distribution \\
            $\mu_a(t,k)$ & Utility received by $a$ with $\vec{D}_a(t)$ over $k$ time units \\
            $\hat{\mu}_a(t-k)$ & Normalized total utility received by $a$ with $\vec{D}_a(t-k)$ \\
            $\bar{\mu}_g(t,S)$ & The global expected normalized utility for strategy $S$ \\
        \bottomrule
    \end{tabular}
    \caption{Notation table.}
    \label{tab:notation_table}
\end{table}

\section{Modeling Strategic Behavior} \label{sec:3}
To identify strategic behavior for content production, we need to address two questions. First, how to determine the strategy distribution $\vec{D}_c$ for content $c$, jointly authored by a set of authors $\mathcal{A}_c$. Notice that each author $a \in \mathcal{A}_c$ has an individual strategy distribution $\vec{D}_a(t)$. Second, we need to determine how the prior strategy distribution $\vec{D}_a(t-1)$ and the strategy distributions of the co-authors of $a$ influence the strategy distribution $\vec{D}_a(t)$.

Next, we introduce key modeling assumptions followed by an elegant Dynamic Dual Attention Network (DDAN) to jointly solve both questions.

\subsection{Assumptions} \label{sec:3.1}
Now we discuss assumptions useful for developing our model.

\begin{description}
\item[Strategy distributions:] We associate a strategy distribution $\vec{D}_c$ with content $c$ produced at time $t$. $\vec{D}_c$ depends on the individual strategy distributions $\vec{D}_a(t)$ of the set of authors $\mathcal{A}_c$ who jointly produce $c$. In other words, the set of authors $\mathcal{A}_c$ draw the strategy $S_i \in \mathcal{S}$ given $\vec{D}_c$ to determine attributes for $c$. Assume that an author $a$ participates in the production of a set of contents $\mathcal{C}_a(t)$. We assume that two factors influence her strategy distribution $\vec{D}_a(t)$: her prior strategy distribution $\vec{D}_a(t-1)$; the strategy distribution of her co-authors for each $c \in \mathcal{C}_a(t)$.

\item[Utility calculation:] Let $\mu_c(k)$ be the utility accumulated by content $c$ after $k$ time units. Since each author $a \in \mathcal{A}_c$ contributes to a different extent to produce $c$, we assume that the utility that flows back to $a$ is in proportion to her contribution. That is, the utility $\mu_{a \mid c}(k) \propto \mu_{c}(k) \cdot r( a \mid c)$, where $\mu_{a \mid c}(k)$ is the utility that flows back to $a$ after $k$ time units in proportion to her contribution $r( a \mid c)$. Notice that $\sum_a r( a \mid c)=1, a \in \mathcal{A}_c$.

\item[Vertex representation:] We associate each content $c$ with a node embedding vector $\vec{h}_c \in \mathbb{R}^{F}$ and each author $a$ at time $t$ with a node embedding vector $\vec{h}_a(t) \in \mathbb{R}^{F}$ (e.g., ESim~\cite{shang2016meta}). We obtain a time-dependent embedding vector for an author, by treating the same author at different times as separate nodes when embedding the network.

\item[Network snapshots:] Since the graph $G = (\mathcal{V},\mathcal{E})$ grows over time, we divide the graph into snapshots. Specifically, we define the vertex set $\mathcal{V}(t)$ and the edge set $\mathcal{E}(t)$ for the graph $G(t)$ to include the authors active at time $t$, the contents created at time $t$, and the links from the content created at time $t$ with their attributes. If an author appears for the first time in snapshot $t$, we draw the prior strategy distributions $\vec{D}_a(t-1)$ from a flat Dirichlet distribution and use an all zero vector as the prior embedding $\vec{h}_a(t-1)$.
\end{description}

\subsection{Dynamic Dual Attention Networks} \label{sec:3.2}
We propose a novel Dynamic Dual Attention Network (DDAN), inspired by the work on Graph Attention Networks by \citet{velivckovic2017graph}, to identify the strategy distributions for content and its authors. The DDAN elegantly addresses the two central dependencies: the strategy for the production of any content depends on the strategies of its authors, and an author's prior strategy as well as her co-authors influence her current strategic behavior. We jointly optimize two attention mechanisms.

\subsubsection{Determining the strategy for the production of a single content} \label{sec:3.2.1}
The strategy distribution $\vec{D}_c$ of a content $c$ created at time $t$ is affected by the strategy distribution $\vec{D}_a(t)$ of all its authors $a \in \mathcal{A}_c$. To determine the contribution $\alpha_{a|c}$ of a specific author $a$ towards $\vec{D}_c$, we feed the embedding vector of the content $c$ (i.e., $\vec{h}_c$) and of the author $a$ at time $t$ (i.e., $\vec{h}_a(t)$) into a one-layer attention mechanism as follows:
\begin{align} \label{eq:con_att_aut}
    e_{a|c} & = \sigma \left( \vec{\phi}_{c,a}^\top \cdot \big[ \mathbf{W}_{c,a} \vec{h}_c \, || \,  \mathbf{W}_{c,a} \vec{h}_a(t) \big] \right), \\
    \alpha_{a|c} & = \mbox{softmax}_a(e_{a|c}) = \frac{\exp( e_{a|c} )}{\sum_{a'\in\mathcal{A}_c}\exp(e_{a'|c})},
\end{align}
where $\mathbf{W}_{c,a} \in \mathbb{R}^{F'\times F}$ is a shared linear transformation and $\vec{\phi}_{c,a} \in \mathbb{R}^{2F'}$ is the weight vector in a one-layer feedforward neural network. Note that $||$ is the concatenation operator to concatenate two vectors and we use \texttt{LeakyReLU} for the nonlinearity $\sigma$. We use \texttt{softmax}  normalization to ensure that the contributions of all the coauthors to a particular content sum to $1$. Finally, note that since $\alpha_{a|c}$ is $a$'s contribution to the determination of $\vec{D}_c$, we set $r( a \mid c) = \alpha_{a|c}$.

Then the strategy distribution $\vec{D}_c$ of content $c$ is the sum of its authors' strategy distributions $\vec{D}_a(t)$ at time $t$, weighted by each authors $a$'s contribution $\alpha_{a|c}$. We use $\xi$ to represent $\tanh$ nonlinear activation. We use $L_1$ normalization to ensure that $\vec{D}_c$ is a valid strategy distribution:
\begin{align} \label{eq:con_dist}
    \vec{D}_c & = \xi \left(\sum_{a \in \mathcal{A}_c} \alpha_{a|c} \cdot \vec{D}_a(t)\right).
\end{align}

\subsubsection{Determining an author's strategy} \label{sec:3.2.2}
An author's strategy $\vec{D}_a(t)$ depends on the strategy adopted for each content she authors at time $t$ as well as her past strategy distribution $\vec{D}_a(t-1)$. First, we examine the effect of the strategy for the production of content $c$ where she is a co-author in $\mathcal{A}_c$.

We apply an attention mechanism to learn content $c$'s contribution $\alpha_{c|a}$ on author $a$'s strategy distribution $\vec{D}_a(t)$ as follows:
\begin{align} \label{eq:aut_att_con}
    e_{c|a} & = \sigma \left( \vec{\phi}_{a,c}^\top \cdot \big[ \mathbf{W}_{a,c} \vec{h}_a(t) \, || \,   \mathbf{W}_{a,c} \vec{h}_c \big] \right), \\
    \alpha_{c|a} & = \mbox{softmax}_c(e_{c|a}) = \frac{\mbox{exp}(e_{c|a})}{\sum_{c'\in\mathcal{C}_a(t)}\mbox{exp}(e_{c'|a})}.
\end{align}

We use a different attention mechanism to determine the contribution of $a$'s strategy distribution at time $t-1$ on her current strategy distribution:
\begin{equation} \label{eq:aut_att_aut}
    \beta_a(t) = \mbox{sigmoid} \left( \vec{\phi}_{a,a}^\top \cdot \big[ \mathbf{W}_{a,a} \vec{h}_a(t) \, || \,  \mathbf{W}_{a,a} \vec{h}_a(t-1) \big] \right).
\end{equation}

Thus, $a$'s strategy distribution $\vec{D}_a(t)$ at time $t$ is the weighted sum of the strategy distribution $\vec{D}_c$ for $c \in \mathcal{C}_a(t)$, and $\vec{D}_a(t-1)$: 
\begin{align} \label{eq:aut_dist}
    \vec{D}_a(t) & = \xi \left( \beta_a(t)\vec{D}_a(t-1) + (1-\beta_a(t))\sum_{c \in\mathcal{C}_a(t)} \alpha_{c|a} \cdot \vec{D}_c \right).
\end{align}
We use $L_1$ normalization to ensure that $\vec{D}_a(t)$ is a valid distribution.

In this section, we discussed modeling assumptions and showed how to determine the content production strategy as a function of the author strategies as well as how past author strategy $\vec{D}_a(t-1)$ and co-author strategies influence an author's current strategy $\vec{D}_a(t)$. Next, we show how to model a rational agent that maximizes expected utility.

\section{A Model for Rational Behavior } \label{sec:4}
 Rational behavior is a useful baseline to understand better the behaviors that we discover in this paper. An author engaged in rational behavior would be able to evaluate the utilities of all strategies and be able to identify the optimal strategy. However, determining rational behavior is hard for several reasons. First, notice that while the actions (e.g., the paper that author cites; content topic) are observable, \textit{the strategies that result in the actions (e.g., pick highly cited papers to cite) are not observable}. This means that any author with access to unlimited resources, who wishes to engage in rational play, will need to develop a model of user behavior that connects strategies to outcomes, fit the model, and then connect strategy distributions $\vec{D}_a$ to payoffs. Unsurprisingly, we could develop several plausible rational models, and below, we discuss one such model that utilizes our DDAN framework.
 
 First, we ask: given the utility at time $t$ of content co-authored by $a$ at time $t-k$, what is the utility of author $a$ using strategies distribution $\vec{D}_a(t-k)$? We compute a normalized utility as follows:
 \begin{equation}
    \hat{\mu}_a(t-k) = \frac{1}{\underbrace{k}_{\text{time}}} \times \frac{\sum_{c \in \mathcal{C}_a(t-k)} \mu_{a|c}(t-k) }{ 
        \underbrace{|\mathcal{C}_a(t-k)|}_{\text{count}}
        },
\end{equation}

The equation says that the normalized utility $\hat{\mu}_a(t-k)$ due to the distribution $\vec{D}_a(t-k)$ depends on the relative utility $\mu_{a|c}(t-k)$ accruing due to participation in the creation of $c \in \mathcal{C}_a(t-k)$. We normalize this sum utilities by the number of content co-authored by $a$ at time $t-k$ and further normalized by the time elapsed $k$, between content production and evaluation.  In our model, to simplify analysis, we allocate the utility $\hat{\mu}_a(t-k)$ to the maximum likelihood strategy in $\vec{D}_a(t-k)$.


A rational author $r$ would thus learn the global (i.e., over all authors) expected utility $\bar{\mu}_g(t,S)$ for each strategy $S \in \mathcal{S}$. Using the expected value allows us to average out over unobserved confounds that may co-vary with utility (e.g., author institution; prior author reputation). Next, we present our experimental results.

\begin{figure*}[ht]
\includegraphics[width=\textwidth]{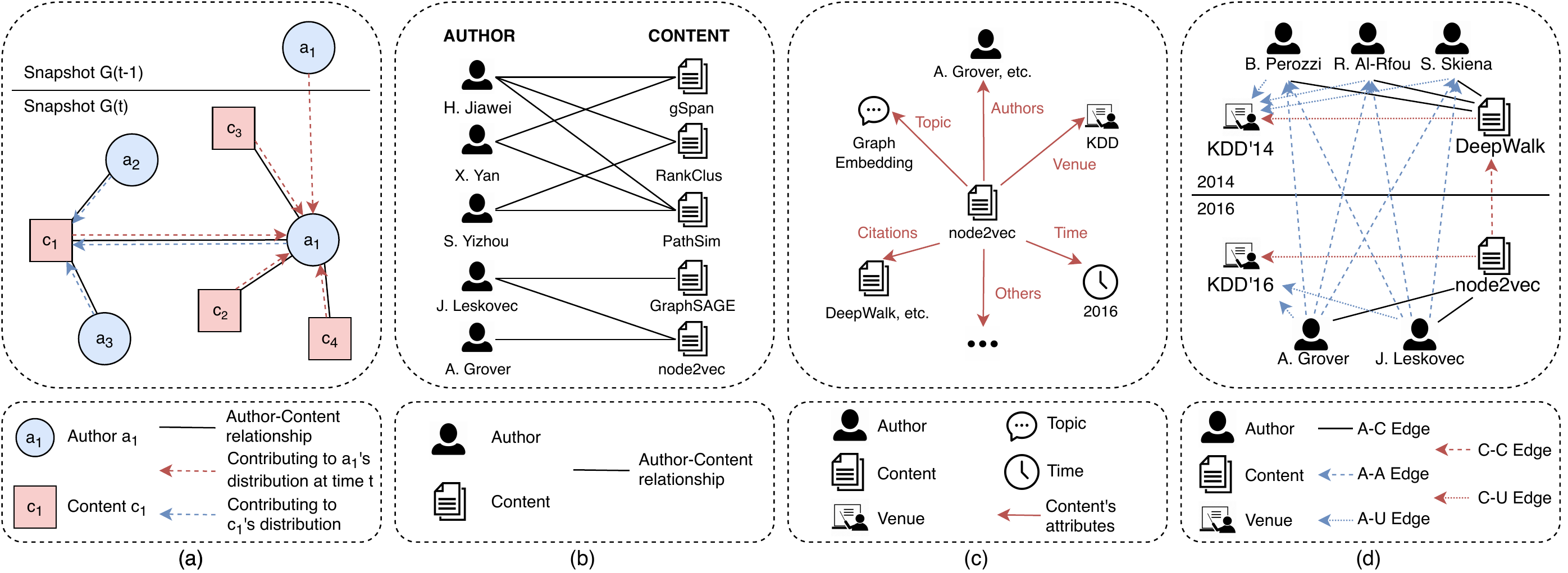}
  \caption{(a) DDAN framework applied to snapshot $G(t)$. (b) An illustration of the author-content bipartite graph based on academic papers. (c) An illustration of the attributes of a content based on academic papers. (d) An example of the social network in~\Cref{sec:5} when we consider two attributes of content: links to prior contents \& publication venues.} \label{fig:section5}
\end{figure*}
\section{Experiments} \label{sec:5}
In this section, we discuss our datasets in~\Cref{sec:5.1}, strategy spaces in~\Cref{sec:5.2}, DDAN training in~\Cref{sec:5.3}, experiment setting in~\Cref{sec:5.4}, competing baselines in~\Cref{sec:5.5} and evaluation in~\Cref{sec:5.6}. We used XSEDE~\cite{xsede} resources for computation and our implementations are publicly available~\footnote{\url{https://github.com/CrowdDynamicsLab/Discovering_Strategic_Behaviors}}.
\subsection{Datasets} \label{sec:5.1}
We use the latest version of the DBLP academic dataset~\cite{Tang:08KDD, sinha2015overview}. The goal with this dataset is to discover strategic behavior associated with two paper attributes: citations and publication venue. That is, \textit{what are the strategic considerations behind whom to cite, and where to publish?} We set papers created during the years 1980--1999 as the background papers. We aim to examine the strategies adopted by authors starting from the year 2000 and use $G(t)$ for $t \in \{1,\dots,19\}$ to represent the status of the network corresponding to each year between $2000$ and $2018$ inclusive. Not every remaining paper in the dataset contains both citation and location information. We infer an author or paper's citation or location strategies only if we can observe the corresponding citation or location edges. This corresponds to 97\% of the papers.
\subsection{Strategy Spaces} \label{sec:5.2}
Now, we discuss four strategy spaces that alter the distributions of the attributes. As we discussed in~\Cref{sec:2.2}, a bipartite graph $G_{a,c}$ represents the content production, connecting authors to the content that they help co-author. One can create using the attributes of each paper, additional graphs: an author-author citation graph $G_{a,a}$, a paper-paper citation graph $G_{c,c}$, a paper-location graph $G_{c,u}$ and an author-location graph $G_{a,u}$. 
Since our DDAN is symmetric with respect to content and authors, in this section we will discuss strategic considerations for content (i.e. focus on explaining $G_{c,c}$ and $G_{c,u}$); similar arguments hold for authors. Thus, consider a paper $c_1$ that cites $c_2$ and is published at location $u_1$. We need to identify strategic considerations that explain the directed edges $(c_1, c_2)$ and $(c_1, u_1)$.
We identify four aspects (see~\Cref{tab:pure_strategies}) based on popularity, similarity of field, familiarity and time recency. As a reminder, each strategy alters the posterior distributions of the attribute value, and \textit{does not} deterministically set the attribute value. When picking papers to cite, authors may pick highly cited papers, from similar fields. They may also pick papers by authors whom they know (e.g. papers by past co-authors), and if the paper topic is in a newly emerging area of research, more recent papers. 
\begin{table}[t]
    \centering
    \begin{tabular}{cl}
        \toprule
            Aspect & Strategy  \\
        \midrule
            \multirow{2}{*}{Popularity} & $s_{1,0}$, preferential attachment \\ 
             & $s_{1,1}$, uniform attachment \\
            \multirow{2}{*}{Field} & $s_{2,0}$, preferring similar fields \\
             & $s_{2,1}$, preferring distinct fields \\
            \multirow{2}{*}{Familiarity} & $s_{3,0}$, preferring familiar nodes \\
            & $s_{3,1}$, preferring unfamiliar nodes \\
            \multirow{2}{*}{Time} & $s_{4,0}$, preferring small time gaps \\
            & $s_{4,1}$, choosing random time gaps \\
        \bottomrule
    \end{tabular}
    \caption{Meaning of each pure strategy. Each composite citation strategy consists of one pure strategy from each of the four aspects. Each composite location strategy consists of one pure strategy from each of the first three aspects.}
    \label{tab:pure_strategies}
\end{table}
\subsubsection{Popularity} \label{sec:5.2.1} 
We use two strategies to explain  the directed edge $(c_1, c_2)$ (and $(c_1, u_1)$) based on popularity. The first, is preferential attachment (i.e., strategy $s_{1,0}$), documented by~\citet{barabasi1999emergence}, where the probability of citing a past paper is proportional to its citations, as the strategy to pick highly cited papers (or publication venues; for example, authors may want to publish in journals with high impact factor). The second is to pick papers (or locations to publish at) uniformly at random (i.e., strategy $s_{1,1}$).
\subsubsection{Field} \label{sec:5.2.2}
We use LSA \cite{deerwester1990indexing} to assign each content a 100-dimension field vector $\vec{F}_c$. 
Then an author's field vector $\vec{F}_a(t)$ at time $t$ is the average of the field vectors of contents that he has created by time $t$. A publication venue's field vector $\vec{F}_u(t)$ at time $t$ is the average of the field vectors of contents published there by time $t$. We perform $L_2$ normalization on all the field vectors.
To support their arguments, authors are likely to cite thematically similar papers (i.e. topic homophily~\cite{McPherson2001,Kossinets2009}), or publish in venues with fields similar to the field of the paper. High-impact papers, on the other hand, often cite papers outside of their field~\cite{Uzzi2013}. Thus, we may explain edge $(c_1, c_2)$  either with homophily (i.e., strategy $s_{2,0}$) or with choosing from different fields (i.e., strategy $s_{2,1}$), if the central theme of $c_1$ spans multiple fields. Thus, we can set the likelihood of edge $(c_1,c_2)$ based on strategy $s_{2,0}$ to be $\ell((c_1, c_2) \mid s_{2,0}) \propto \exp(|| - \vec{F}_{c_1} - \vec{F}_{c_2}||)$. The likelihood of choosing a paper $c_2$ from a field distinct from $c_1$ is just the complement of $\ell((c_1, c_2) \mid s_{2,0})$. We make similar arguments for explaining edge $(c_1, u_1)$.
\subsubsection{Familiarity} \label{sec:5.2.3}
Some papers may preferentially cite other papers based on authorship; for example, self-citation is a well known strategy to boost the popular $h$-index~\cite{Glanzel2006, Engqvist2008}. Thus, we can partition the set of papers published before time $t$ into two disjoint sets: one set $\bm{A}$ that contains papers, each of which has one of the co-authors of $c_1$ as a co-author. The second set $\bm{B}$ is the complementary set, containing papers whose authors do not include any of the co-authors of paper $c_1$. We can make a parallel argument to partition the set of past publication venues. Thus in our first familiarity based strategy $s_{3,0}$, a paper $c_1$ will cite another $c_2$ with a high probability if $c_2 \in \bm{A}$ and with a low probability if $c_2 \in \bm{B}$. The converse is true for strategy $s_{3,1}$. 
\subsubsection{Time} \label{sec:5.2.4}
Paper citations also exhibit recency bias~\cite{ghosh2011time}, and thus time is an important factor for explaining edge $(c_1, c_2)$. Since content created at $t$ cannot occur at venues active \textit{prior to} time $t$, we do not include time as a strategic consideration for selecting venue, that is, to explain edge $(c_1, u_1)$. To incorporate recency bias (i.e., strategy $s_{4,0}$), we do the following. Assume that the normalized time difference between the publications $c_1$ and $c_2$ is $0 \leq \delta \leq 1$. Then, to model recency bias, we use a Beta distribution to alter the posterior probability of selecting papers to cite. That is, likelihood $\ell((c_1, c_2) \mid s_{4,0}) \propto B(1-\delta \mid \alpha, \beta)$, where $\alpha, \beta$ are parameters of the Beta distribution. To model recency, we set $\alpha=10, \beta=1$. In the complementary strategy (i.e., strategy $s_{4,1}$), we pick a paper uniformly at random with respect to time of publication.
\subsubsection{Composite Strategies} \label{5.2.5}
Thus far, we discussed four different strategic considerations to explain edge $(c_1, c_2)$: popularity, field, familiarity and time recency. We identify three strategic considerations to explain edge $(c_1, u_1)$: popularity, field, familiarity. Thus the likelihood of the edge $(c_1, c_2)$ is a composite of each of the four strategies. Since each strategic consideration has two possibilities, we can enumerate $2^4=16$ composite strategies to explain edge $(c_1, c_2)$. Correspondingly, we can enumerate $2^3=8$ composite strategies to explain edge $(c_1, u_1)$.
A composite citation strategy consists of a pure strategy under each of the four aspects (Popularity, Field, Familiarity, Time); a composite location strategy consists of a pure strategy under each of the first three aspects (Popularity, Field, Familiarity). For easy reference, we use a binary sequence to represent composite strategies w.r.t. pure strategies (e.g., citation strategy $S_4^c=s_{1,0}\times s_{2,0}\times s_{3,1}\times s_{4,0}$, location strategy $S_6^l=s_{1,0}\times s_{2,1}\times s_{3,1}$). Then the likelihood of forming an edge $e=(c_1,c_2)$ given a composite strategy $S_i$ is the product of the likelihoods of forming that edge given each of $S_i$'s constituent pure strategies.
In this subsection, we explained the strategic considerations to help explain the formation of edges $(c_1, c_2)$ and $(c_1, u_1)$. We can use the same strategies to help explain the formation of edges $(a_1, a_2)$ (i.e., author $a_1$ cites author $a_2$) and $(a_1, u_1)$ (i.e., author $a_1$ publishes in venue $u_1$)

\subsection{DDAN Training \& Optimization} \label{sec:5.3}
Now, we we discuss how to train and optimize the Dynamic Dual Attention Networks (DDAN). We first initialize $\vec{D}_c$ and $\vec{D}_a(t)$ as generated by the flat Dirichlet distribution for content set and authors active at time $t$. Then we train the attention networks to update $\vec{D}_c$ and $\vec{D}_a(t)$ alternatively by using the current snapshot $G(t)$ as the ground truth. The overall framework is illustrated in~\Cref{fig:section5}. 
We need to explain four graphs using DDAN: an author-author citation graph $G_{a,a}$, a paper-paper citation graph $G_{c,c}$, a paper-location graph $G_{c,u}$, and an author-location graph $G_{a,u}$. As a concrete example, consider the graph $G_{c,c}$. Then, at time $t$, we need to minimize the negative log likelihood:
\begin{equation} \label{eq:3.5_loss_func}
    L_{c,c}(t) = \sum_{(c_i,c_j)\in \mathcal{E}_{c,c}(t)} -\log \sum_{S_i \in \mathcal{S}} P(S_i \mid D_{c_i}) \cdot \ell((c_i,c_j) \mid S_i)
\end{equation}
Where, $P(S_i \mid D_{c_i})$ is the probability of picking strategy $S_i$ given the distribution $\vec{D}_{c_i}$ for content $c_i$, and $\ell((c_i,c_j) \mid S_i)$ is the likelihood of edge $(c_i, c_j)$ given strategy $S_i$.  Thus~\Cref{eq:3.5_loss_func} states that we need to sum over all edges $(c_i, c_j)$, the negative log of the likelihood of observing edge $(c_i, c_j)$ conditioned on strategy distribution $\vec{D}_{c_i}$. We can construct similar likelihood functions to explain graphs $G_{a,a}$, $G_{c,u}$ and $G_{a,u}$. The overall likelihood is just a sum of the constituent likelihoods. That is, $L(t) = L_{c,c}(t) + L_{c,u}(t) + L_{a,a}(t)+L_{a,u}(t)$. Once the DDAN converges in the current snapshot $G(t)$, we move on to the next snapshot $G(t+1)$ until all the snapshots are covered.
\subsection{Experiment Settings} \label{sec:5.4}
To properly evaluate the strategy distributions identified by the proposed DDAN framework, we apply them to the task of link prediction. We model the link prediction problem as a recommendation problem which aims to rank node pairs in terms of the posterior probability of forming an edge between them. That is, the identified strategy distributions should best explain the observed network.
We identify the set of authors $\widetilde{\mathcal{A}}'(t)$ with over five new contents in the current snapshot $G(t)$ and partition each author's contents for 5-fold cross validation. For each fold $\mathcal{C}_a(t,j)$ where $j \in \{1,\dots,5\}$, we hide the author-content edges between $a$ and $c \in \mathcal{C}_a(t,j)$ as well as the edges between $a$ and attribute nodes $\mathcal{A}^+_a(t,j)$ and $\mathcal{U}^+_a(t,j)$ formed due to $c \in \mathcal{C}_a(t,j)$. The model is trained using the remaining network and aims to recover the hidden attribute edges. Since exhaustive computation over all node pairs is expensive, we utilize the information of $a$'s coauthors when creating $c \in \mathcal{C}_a(t,j)$. The authors cited by those co-authors and the venues where those coauthors have made publications by time $t$ constitute the negative testing sets $\mathcal{A}^-_a(t,j)$ and $\mathcal{U}^-_a(t,j)$, respectively. 
We apply this process on the DBLP dataset with a five-year gap (i.e., we only look at snapshots corresponding to Year 2000, 2005, 2010, 2015, and 2018). We summarize these statistics in~\Cref{tab:testset_statistics}. We use Mean Average Precision (MAP) as the evaluation metric.
\begin{table}[t]
    \centering
    \begin{tabular}{@{}crcc@{}}
        \toprule
            Snapshot Year & {$|\widetilde{\mathcal{A}}'(t)|$} & {$|\mathcal{A}_a^+(t,j)|$} & {$|\mathcal{U}_a^+(t,j)|$} \\
        \midrule
            2000 & 3,145 & 31.48 & 1.34 \\
            2005 & 10,316 & 42.34 & 1.47 \\
            2010 & 18,062 & 57.11 & 1.50 \\
            2015 & 25,759 & 82.18 & 1.56 \\
            2018 & 9,192 & 109.64 & 1.41 \\
        \bottomrule
    \end{tabular}
    \caption{Statistics of the test set. $|\widetilde{\mathcal{A}}'(t)|$ is the size of the set of authors with over five new contents at time $t$. $|\mathcal{A}_a^+(t,j)|$ and $|\mathcal{U}_a^+(t,j)|$ are the average sizes of the positive testing sets per author per fold at time $t$ when we examine the citation and publication strategies, respectively.}
    \label{tab:testset_statistics}
\end{table}
\subsection{Baselines} \label{sec:5.5}
We want to point out that the problem of identifying authors' strategy distributions can also be modeled by topic generative models. Therefore, we compare our DDAN framework against two different topic models \cite{yin2014dirichlet, wang2006topics} and one traditional regression model \cite{hosmer2013applied}. For static models \cite{yin2014dirichlet, hosmer2013applied}, we apply the model to each testing snapshot separately. For dynamic models \cite{wang2006topics} and DDAN, we first obtain the history information with the entire dataset, and then apply the model to the training sets in each testing snapshot.
\begin{enumerate}
    \item \textbf{Logistic Regression (LR)} \cite{hosmer2013applied}: When applying the logistic regression model on each author individually, we treat the likelihood of forming an edge as the predictor and the ground truth as the response variable. The coefficients in the regression model are constrained to be non-negative and sum to 1 so that they can be interpreted as strategy distributions. 
    \item \textbf{Dirichlet Multinomial Mixture Model (DMM)} \cite{yin2014dirichlet}: Words become strategies and topics over words become distributions over strategies. Authors need to pick one strategy from their strategy distributions to form an edge. We set the number of topics to be the same as the number of strategies so that each topic is initialized with a maximum likelihood strategy. All the authors in each testing snapshot are trained together to detect their strategy distributions.
    \item \textbf{Topics Over Time (TOT)} \cite{wang2006topics}: In comparison with DMM, TOT requires authors to first choose a topic from her distribution over topics in that snapshot and then pick a strategy from the chosen topic to form an edge. Meanwhile, each topic is also associated with a continuous distribution over time snapshots.
    \item \textbf{DDAN}: The proposed framework which models individuals' content production strategies under the influence of social interactions involved in the process.
\end{enumerate}
\begin{table}[b]
    \centering
    \begin{tabular}{@{}cccccc@{}}
        \toprule
            Year & Strategies & LR~\cite{hosmer2013applied} & DMM~\cite{yin2014dirichlet} & TOT~\cite{wang2006topics} & DDAN \\
        \midrule
            \multirow{2}{*}{2000} & Citation & 0.72 & 0.72 & 0.73 & \textbf{0.74} \\
            & Publication & 0.71 & 0.70 & 0.73 & \textbf{0.75} \\
            \multirow{2}{*}{2005} & Citation & 0.69 & 0.69 & 0.70 & \textbf{0.71} \\
            & Publication & 0.69 & 0.69 & 0.72 & \textbf{0.73} \\
            \multirow{2}{*}{2010} & Citation & 0.67 & 0.67 & 0.68 & \textbf{0.69} \\
            & Publication & 0.71 & 0.71 & 0.73 & \textbf{0.74} \\
            \multirow{2}{*}{2015} & Citation & 0.67 & 0.67 & 0.68 & \textbf{0.69} \\
            & Publication & 0.72 & 0.72 & 0.74 & \textbf{0.75} \\
            \multirow{2}{*}{2018} & Citation & 0.67 & 0.67 & 0.68 & \textbf{0.69} \\
            & Publication & 0.76 & 0.75 & 0.77 & \textbf{0.78} \\
        \bottomrule
    \end{tabular}
    \caption{Experiment results using Mean Average Precision (MAP) as the evaluation metrics. DDAN achieves the highest scores for both strategies in all testing snapshots.}
    \label{tab:experiment_results_table}
\end{table}
\subsection{Evaluation} \label{sec:5.6}
Each model is tested using 5-fold cross-validation, and the average MAP scores are reported in~\Cref{tab:experiment_results_table}. We can see that DDAN outperforms the rest for both strategies in all snapshots. LR performs slightly better than DMM since LR is trained on each author individually, while the training of DMM requires the entire set of active authors and may involve some noises. Meanwhile, dynamic models (DDAN, TOT) also give better results than static models (LR, DMM) since dynamic models consider the connection across snapshots. In general, unlike the baseline models (LR, DMM, TOT) which focus on authors' information and ignore the social context, the proposed DDAN framework comprehensively models the interactions between authors and contents within the same snapshot, as well as the dependencies between the present and the past. Therefore, DDAN achieves the highest performance. 
In this section, we introduced four strategic considerations---popularity, field, familiarity, and time---to explain the existence of an edge in four different derived graphs from the DBLP dataset. We also covered DDAN training and reported experimental results comparing DDAN with state-of-the-art baselines. The goal of these baselines was to understand if DDAN modeled the observations well. Next, we present a qualitative analysis of our results.

\begin{figure*}[ht]
\begin{subfigure}{.32\textwidth}
  \centering
  \includegraphics[width=\linewidth]{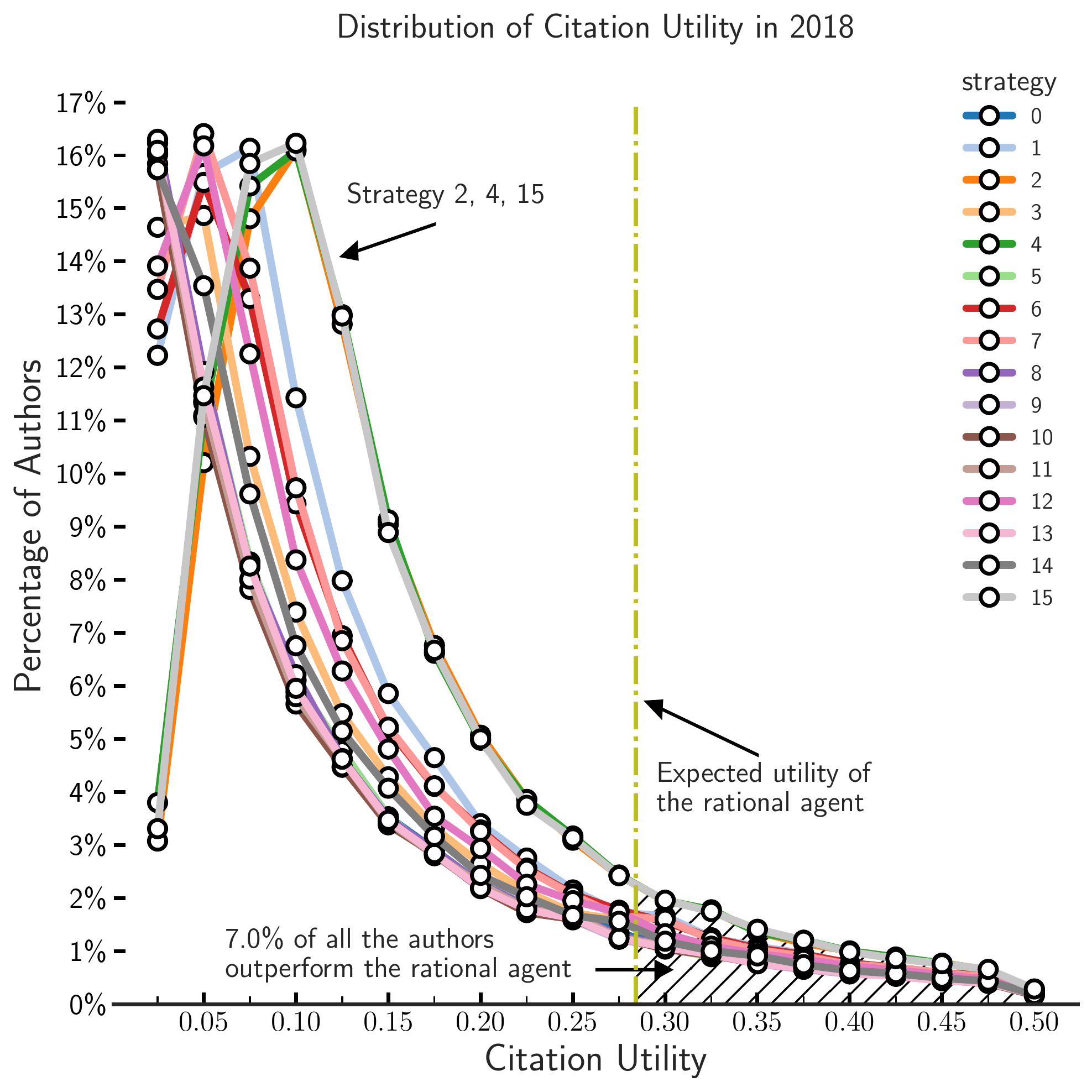}  
  \caption{}
  \label{fig:sub-first}
\end{subfigure}
\begin{subfigure}{.32\textwidth}
  \centering
  \includegraphics[width=\linewidth]{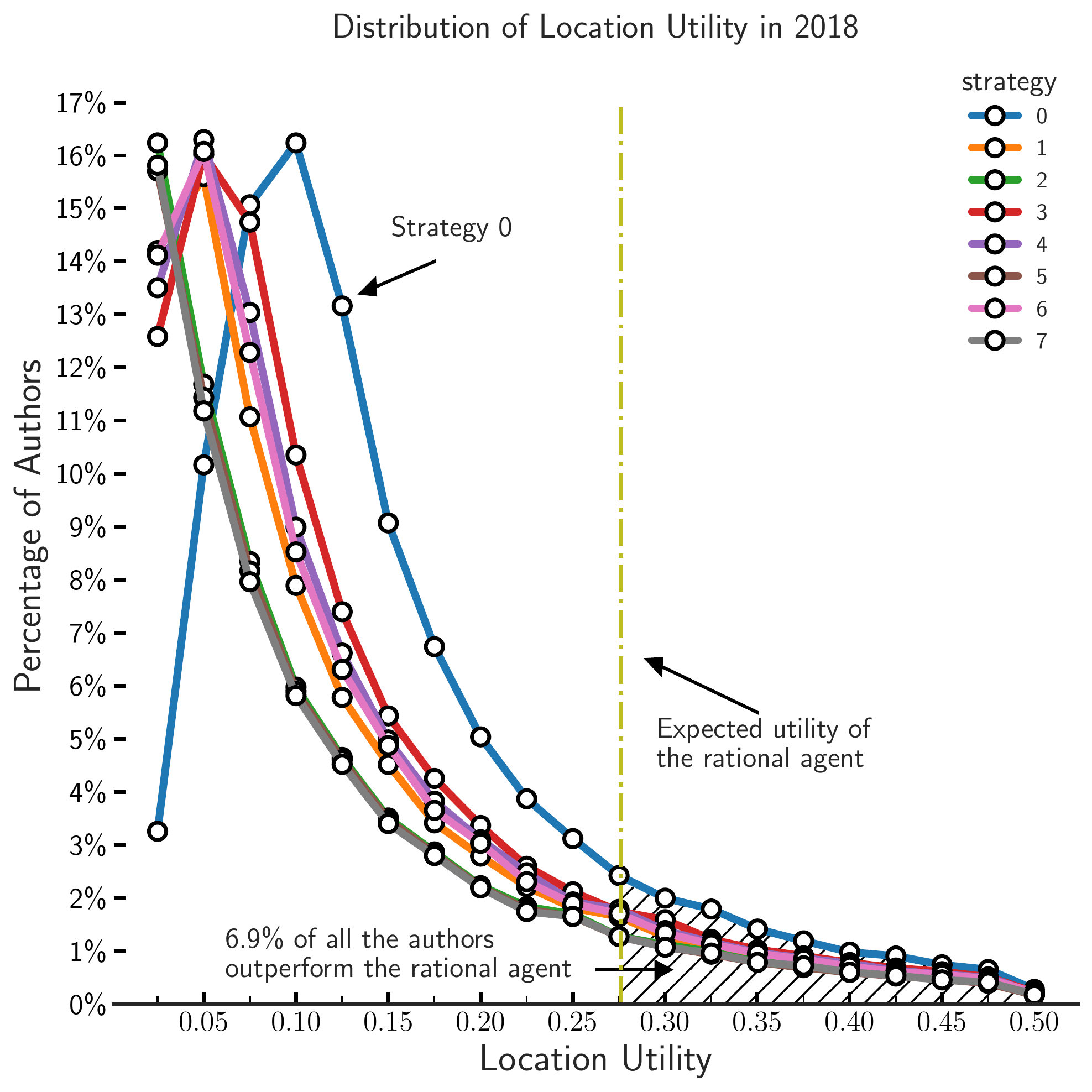}  
  \caption{}
  \label{fig:sub-first}
\end{subfigure}
\begin{subfigure}{.32\textwidth}
  \centering
  \includegraphics[width=\linewidth]{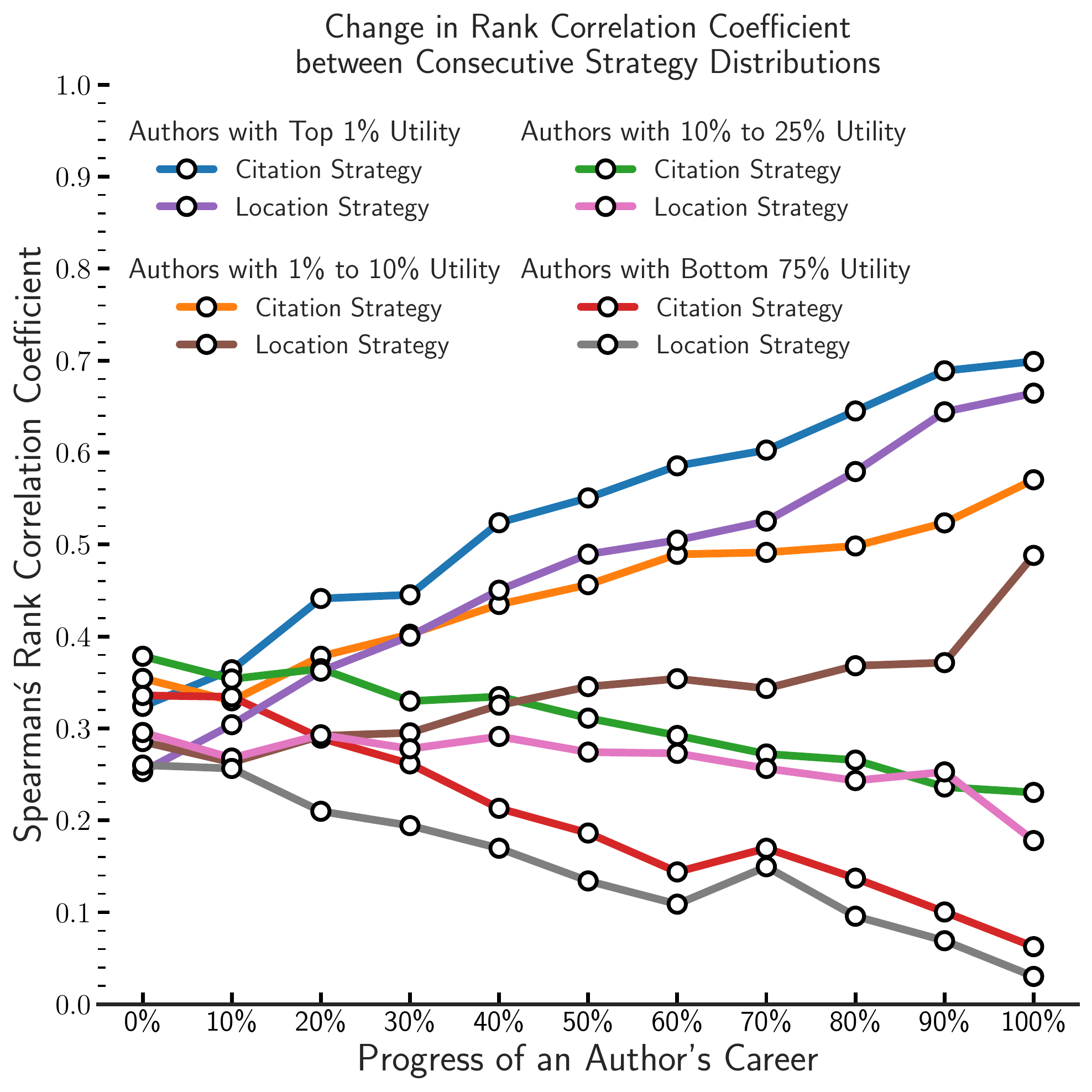}  
  \caption{}
  \label{fig:sub-first}
\end{subfigure}
 \caption{The figures show the distribution of utilities for the citation and location strategy spaces (i.e., (a) and (b)) and the plot of strategy preference order correlation as a function of where authors are in their career (i.e., (c)). Figures (a) and (b) show that different strategies yield different outcomes and that some authors ($\sim 7\%$) do better than the rational agent who maximizes expected utility. Figure (c) shows that only the top $10\%$ of the authors' preference order begins to stabilize throughout their career. A majority ($\sim 75\%$) of authors do not stabilize in their strategy preference order.} \label{fig:combined_curves}
\end{figure*}
\section{Qualitative Analysis} \label{sec:6}
Two questions in~\Cref{sec:2.2} motivated us: if the preference order amongst strategies stabilizes for individuals (thus indicating the emergence of strategic behavior) and if these stable preference orders are correlated with high utility. As a reminder, stable preference orders may not correlate to high payoffs---since individuals are resource-limited, they may lack the resources to discover the correlation between behavior and payoff. Instead, stability may arise due to other factors, including social norms. Let us examine each question in turn.
In the analysis that follows, we use the same strategy spaces and payoffs introduced in the previous section: the authors in an academic social network make decisions on whom to cite, and where to publish. We used sixteen citation strategies and eight location strategies. Since we use the academic dataset, we use the number of citations as the content utility, that is, $\mu_c(k)$ refers to the number of citations $c$ received over $k$ time units after publication. We assume for this analysis that citation strategies (i.e., which papers to cite) and location strategies (i.e., where to publish) are independent and contribute equally to the content utility $\mu_c(k)$.
\subsection{Do strategies matter?} \label{sec:6.1}
Let us first examine if there are any differences amongst the strategies $S \in \mathcal{S}$. Consider~\Cref{fig:combined_curves} (a)-(b). The two figures show the distribution of maximum likelihood utilities over the citation and location strategy space for the whole population. Observe that the utility curves in each sub-figure are distinct: each strategy distribution has a different mode, and some strategies have a higher payoff. 
Consider~\Cref{fig:combined_curves} (b), the distribution of location strategies. It shows that location strategy 0, i.e., $S^l_0$, has the maximum modal payoff (i.e., the distribution with the highest mode). This strategy says that the authors pick venues based on preferential attachment (that is, pick venues in proportion to their publication popularity), that are from similar fields as the author, and that are familiar (i.e., the author has published there earlier).  In hindsight, this is intuitive---by publishing in popular venues, there is an increasing likelihood that their papers will be visible, with similar fields, there is an increasing chance that the paper is more likely to be accepted due to topical match, and if the author has published there previously, then the author understands the social norms in terms of how to write for that audience, again increasing the chance of acceptance.
\Cref{fig:combined_curves} (a), the distribution of citation strategies offers similar insights. It shows that citation strategy 2, 4 and 15, i.e. $S^c_2$, $S^c_4$ and $S^c_{15}$, have the highest modal utilities. Let us examine strategy 4, i.e. $S^c_4$, in detail; the insights for the other two follow a similar argument. $S^c_4$ says that authors cite papers based on preferential attachment (i.e., they cite highly cited papers), from similar fields (i.e., they cite papers similar to their own paper), pick papers that are not familiar (that is, they don't cite their own papers), and pick most recent papers. This strategy of picking papers that are well cited in their own field makes sense---it is less likely that their paper will be rejected for inadequate references; that they don't self-cite is also reasonable since excessive self-citation is frowned upon, and citing more recent papers implies that they have covered all the recent, relevant works in their area. 
\subsection{Emergence of Order} \label{sec:6.2}
Having established that the strategies have different payoffs, let us examine if authors begin to converge on a preferential order over strategies. In our DDAN formulation, each author $a$ has a strategy distribution $D_a(t)$, for each time $t$. We compute the preference order, by utilizing the likelihood of the strategy, for the citation and location strategies for each author. Then, we compute the Spearman rank correlation coefficient \cite{myers2013research}, to compute the correlation in preference order across consecutive publication years, for all authors with at least five publication years. Then, to aggregate across authors, we group rank correlation coefficients in relation to the author career length.
\begin{figure}[t]
\includegraphics[width=\columnwidth]{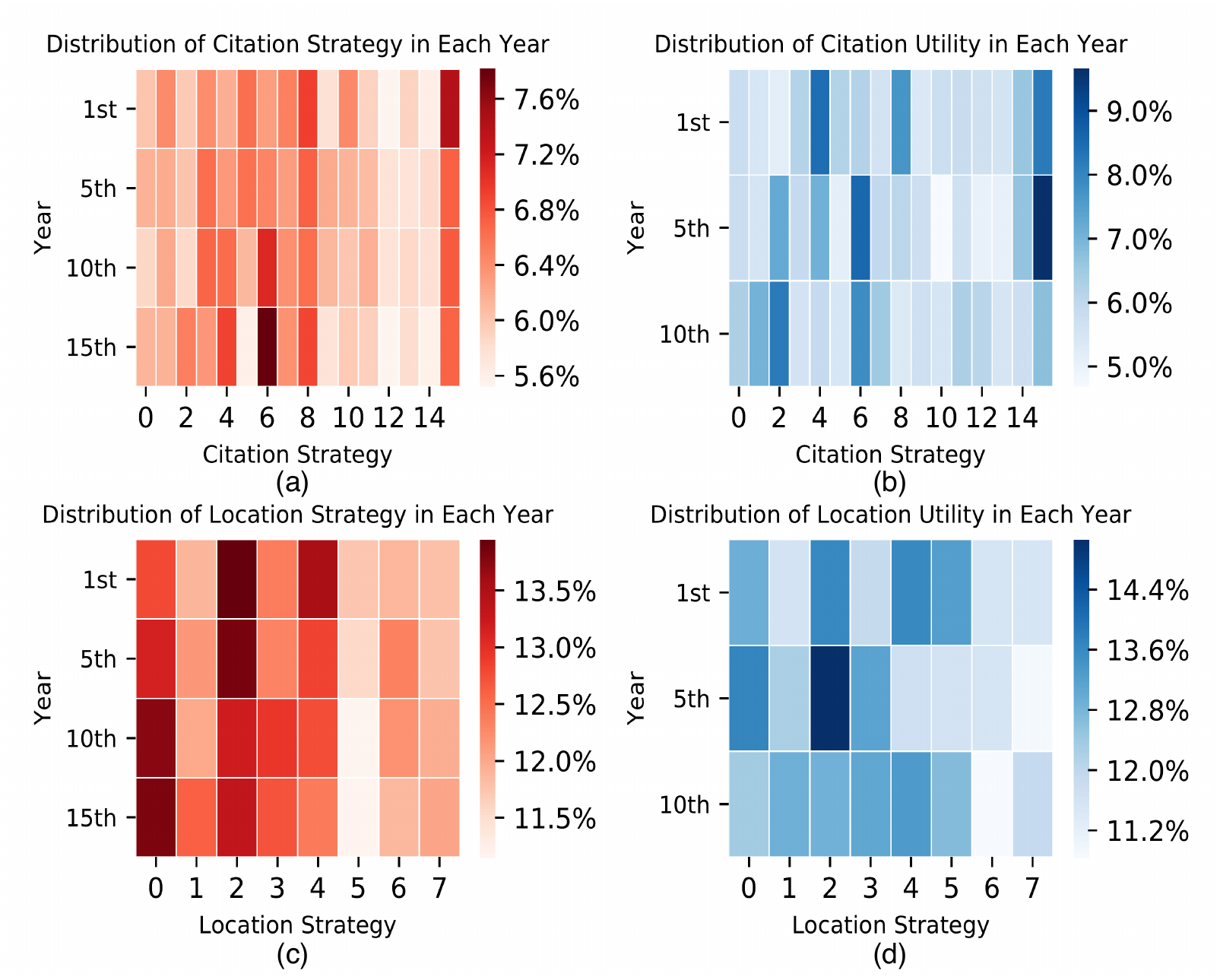}
  \caption{For authors who published in at least 15 years, Figure (a) shows the distribution of their citation strategies in the 1st, 5th, 10th and 15th year. Figure (b) is the corresponding distribution of their citation utilities. Similar heatmaps are drawn for location strategies in Figure (c) and (d).} \label{fig:author_utility_over_time}
\end{figure}
~\Cref{fig:combined_curves} (c) shows the rank correlation curves. The curves show that the correlations \textit{increase} for those authors with the normalized utility in the top $10\%$, with the highest increases for those in the top $1\%$. What the curves imply for the top $1\%$ is that this group quickly converge onto the citation and location strategy, while for the group in the top $1-10\%$ converge onto the citation strategy (but less quickly than do the top $1\%$), but takes a while for them to figure out \textit{where to publish}. What is of note: authors in the bottom $90\%$ in terms of the normalized utility are less likely to be correlated in terms of their citation or location strategy.
~\Cref{fig:author_utility_over_time} shows the strategy distributions for both citation and location strategy spaces, over time, for authors who have published in at least $15$ years (there are 23,238 of them). Notice that over time, for this group of authors, one can see that citation strategy 2, 4, 6 and 15, i.e., $S^c_2$, $S^c_4$, $S^c_6$ and $S^c_{15}$, are beginning to stabilize and location strategy 0, i.e., $S^l_0$, is stabilizing; notice that the corresponding utilities are also high; these plots indicate that for many of these authors, a preference order emerges.
Notice in~\Cref{fig:combined_curves} (a)-(b), we find somewhat surprisingly, that a small percentage of authors ($\sim7\%$) have the normalized utility greater than the rational agent. This is reasonable because the rational agent computes the \textit{expected payoff} of any strategy; there will be some authors for whom that a strategy works better than average. In other words, some authors appear better than the rational agent, because most authors never figure out the correct strategy, depressing the mean.
\subsection{Stability and Payoffs} \label{sec:6.3}
\begin{figure}[t]
  \includegraphics[width=\columnwidth]{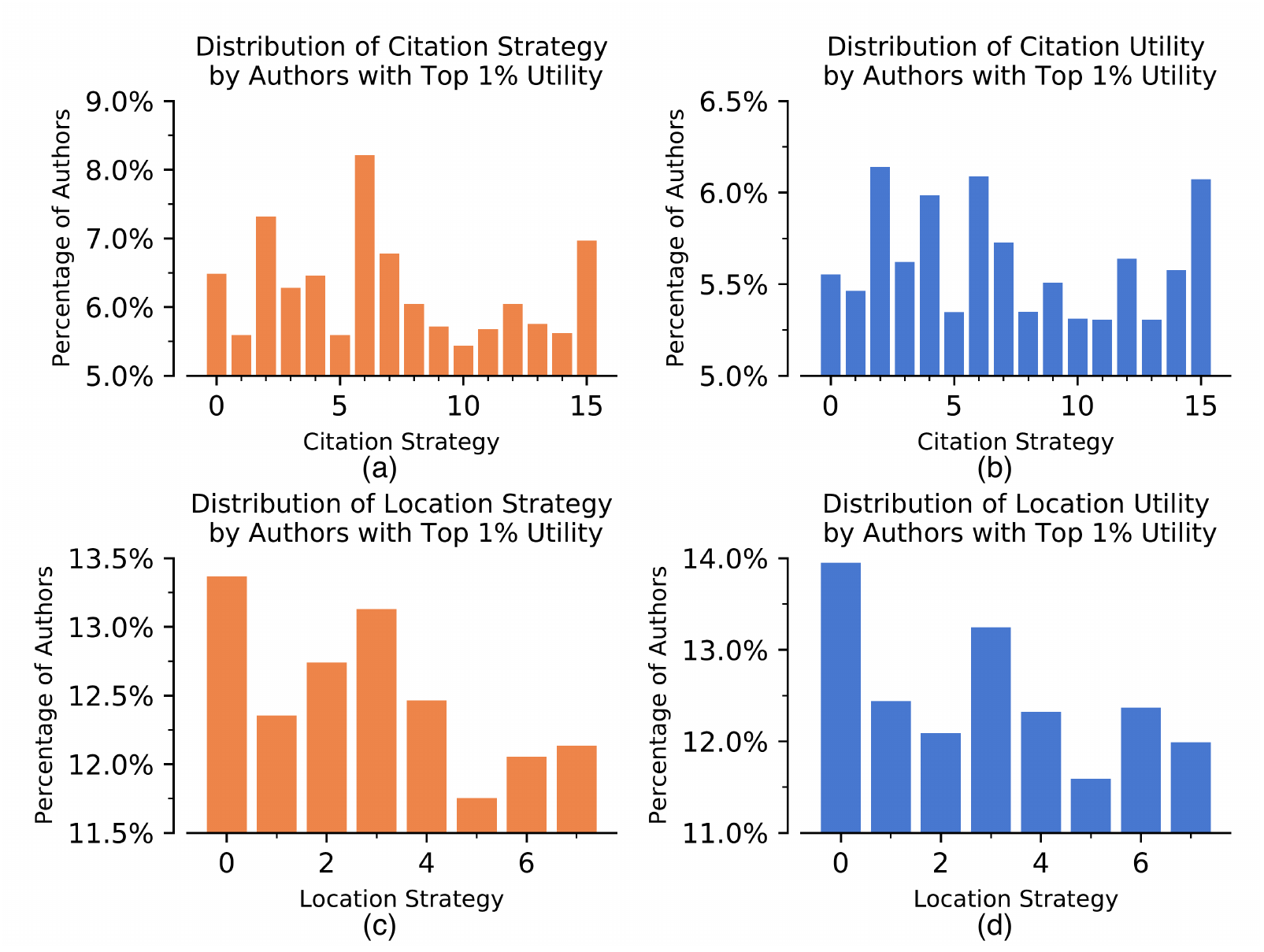}
    \caption{For authors with top $1\%$ normalized utilities, we show the distribution over their citation strategies in Figure (a) and over citation utilities in Figure (b). Similar distributions are drawn for location strategies in Figure (c) and (d). High utility of citation strategy 6 in 5(a) and 5(b) is notable: this strategy emphasizes citing papers from different fields. This is consistent with findings by~\citet{Uzzi2013} that high-impact papers tend to cite papers from different fields.} \label{fig:top1}
  \end{figure}
  
Having established the emergence of a stable preference order for some of the authors, it is natural to ask if this order is correlated with expected utilities. That is, do the preferences over strategies match the utilities that accrue from using them? 
We examine this issue in~\Cref{fig:top1}. The plots show the marginal strategy distribution for the top $1\%$ of the authors in terms of their normalized utility, with the top-left sub-figure showing the marginal citation strategy distribution, and the top right sub-figure showing the corresponding utility distribution. Notice that again, citation strategy 2, 4, 6 and 15, i.e. $S^c_2$, $S^c_4$, $S^c_6$ and $S^c_{15}$, have the highest utility values. While strategy 6 does not have the same high mode as strategy 2, 4 and 15 from~\Cref{fig:combined_curves}, those authors at the top $1\%$ derive more utility from it than most other authors. The main difference between strategy 4 and strategy 6: strategy 6, consistent with work by~\citet{Uzzi2013} on publications with high-impact, emphasizes citing highly cited papers from \textit{different fields}. At the same time, strategy 4 suggests that we pick highly cited papers from the same field. 
A similar pattern emerges for location strategy---strategy 0 being the strategy with the highest payoff for the top $1\%$ is consistent with the findings from~\Cref{fig:combined_curves} (b).
In this section, we qualitatively examined the questions that motivated this paper: the emergence of preference order, and the relationship between preference order and utility. In the case of the academic dataset, and for the citation and location strategy spaces, we observe the emergence of order. Interestingly, we see that preference order is stable only for the top $10\%$ of the individuals in our dataset. Furthermore, we find that the preferences of the top $1\%$ are correlated with utility.

\section{Discussion}
\label{sec:discussion}
In this section, we discuss key ideas and limitations. Below we discuss critical comments: model generalizability, strategy space, and use of one social network dataset.

\begin{description}
    \item[Types of content-production]: Our framework applies to content production scenarios with discrete strategy spaces. While the framework applies to the case when there is a mix of single-authored and collaboratively-authored documents, the proposed dual attention networks make the most sense in the case when the content has multiple authors, for example in venues such as academic citation networks and online forums such as Piazza.
    \item[Generalization to other networks:] Each social network has a different strategy space (the set of possible actions, reward mechanism---mapping actions to utility), social norms, and the network structure (bipartite person and content graph). Thus, interpreting the results of our framework on each social network requires care. This paper discusses results only from the DBLP dataset. We do have results from StackExchange data, where we assumed that the participants collaborated in creating the set of answers, with user reputation as the payoff. The strategy space for StackExchange included, among others, what questions to answer, which users to follow. 
    
    Since this is the first paper on discovering strategic behaviors from networks, we felt that it was essential to perform a qualitative analysis of the discovered behaviors~(\Cref{sec:6}), and not restrict ourselves to a link prediction task~(\Cref{sec:5.6},~\Cref{tab:experiment_results_table}). It is the qualitative analysis that provides a more in-depth understanding (emergence of order; connections between order and payoffs) of behavior. We lacked space in the paper to describe the strategy space for StackExchange (different from DBLP)  \textit{and} perform a careful qualitative analysis of the StackExchange results. We plan to report these results in an extended ArXiv paper. 
    \item[Limitations:] We state two limitations here. First, for each social network, we require a specific set of strategies and a utility function. Identification of a complete strategy space may be non-trivial for some networks. Second, our rational model is myopic; instead, we could use an explore-exploit strategy in the vein of reinforcement learning literature. 
\end{description}

\section{Related Work} \label{sec:7}
Our approach targets strategic behavior modeling in social networks via dual graph attention. We provide a brief overview of related past work.

\textbf{Strategic Behavior Modeling in Social Networks.}
Social behavior modeling is connected to a wide range of past work: \citet{papagelis2011individual} investigates how individual behavior is affected by those of her friends; \citet{kohli2012colonel} looks at how social relations affect players' strategies in a resource allocation game. Others target specific scenarios: \citet{xu2012modeling} studies users' posting behavior on Twitter; \citet{irfan2018power} includes behavioral context in its model of congressional voting; and more recent work models relation types~\cite{rase, igcn} and social influence~\cite{infvae, socialgan} in neural recommendation frameworks; \citet{dong2014inferring} discovers social strategies among mobile users. Some general frameworks are also proposed: \citet{mueller2013general} formally characterizes rational learning in social networks. In contrast to past work, we examine if strategic behaviors emerge among social content producers to maximize social rewards.

\textbf{Graph Attention Networks.}
With recent advancements in graph neural networks \cite{zhou2018graph, wu2019comprehensive, zhang2018deep} and attention mechanisms \cite{vaswani2017attention, lee2018attention}, GAT \cite{velivckovic2017graph} introduces the attention mechanism in network feature aggregation by implicitly prioritizing node neighbors. Several works attempt to extend GAT to dynamic versions: combining recurrent neural networks with GAT~\cite{song2019session};  combined attention on structural neighborhood and temporal dynamics~\cite{sankar2018dynamic} and node-aware attention for user interaction predictions in real-world dynamic graphs~\cite{mlg2019_45}. \citet{wu2019dual} uses user-specific and dynamic context-aware attention weights for social recommendation. In contrast, our proposed DDAN framework extracts latent strategies via forward-backward dual attention to model the social interactions centered on content and authors.

\section{Conclusion} \label{sec:8}
In this paper, we investigated the question of whether resource-limited individuals were able to discover strategic behaviors associated with high payoffs when producing content in social networks. Empirical analysis is challenging since while we do observe the action, and the payoff, we do not observe the strategic considerations underlying the action. Our technical insight was to conceptualize the observed behavior as a generative process. That is, a strategic decision changes the posterior distribution over the action space. We assumed that while the set of strategies is common to all, each individual randomizes over the set of strategies. We proposed a novel framework of Dynamic Dual Attention Networks (DDAN) to model authors' strategic behaviors when creating contents in social networks. We made three strong empirical findings: first, different strategies lead to different payoffs; second, the group of authors with the highest $10\%$ normalized utility exhibit stability in their preferential orders over strategies, which indicates the emergence of strategic behaviors; third, the stability of preference is related to high payoffs.  While our technical insight is generalizable, adapting our framework to other social networks requires care: the strategy spaces of these networks may differ, and if the content-production is not collaborative, DDAN may not be the best approach.

\clearpage
\bibliography{reference.bib}
\bibliographystyle{ACM-Reference-Format}

\end{document}